\documentclass[twocolumn,showpacs,preprintnumbers,amsmath,amssymb,prl]{revtex4}

\usepackage{graphicx}
\usepackage{dcolumn}

\usepackage[latin1]{inputenc}
\usepackage[T1]{fontenc}
\usepackage{ae,aecompl}

\begin{document}

\newcommand{\rar}{$\rightarrow$}
\newcommand{\lrar}{$\leftrightarrow$}

\newcommand{\beq}{\begin{equation}}
\newcommand{\eeq}{\end{equation}}
\newcommand{\bea}{\begin{eqnarray}}
\newcommand{\eea}{\end{eqnarray}}
\newcommand{\Req}[1]{Eq. (\ref{E#1})}
\newcommand{\req}[1]{(\ref{E#1})}
\newcommand{\degree}{$^{\rm\circ} $}
\newcommand{\pcite}{\protect\cite}
\newcommand{\pref}{\protect\ref}
\newcommand{\Rfg}[1]{Fig. \ref{F#1}}
\newcommand{\rfg}[1]{\ref{F#1}}
\newcommand{\Rtb}[1]{Table \ref{T#1}}
\newcommand{\rtb}[1]{\ref{T#1}}

\title{The torque transfer coefficient in DNA under torsional stress}

\author{Alexey K. Mazur}
\email{alexey@ibpc.fr}
\affiliation{UPR9080 CNRS, Université Paris Diderot, Sorbonne Paris Cité,\\
Institut de Biologie Physico-Chimique,\\
13, rue Pierre et Marie Curie, Paris,75005, France.}
 


\begin{abstract}
In recent years, significant progress in understanding the properties
of supercoiled DNA has been obtained due to nanotechniques that made
stretching and twisting of single molecules possible. Quantitative
interpretation of such experiments requires accurate knowledge of torques
inside manipulated DNA. This paper argues that it is not possible to
transfer the entire magnitudes of external torques to the twisting stress
of the double helix, and that a reducing torque transfer coefficient
(TTC<1) should always be assumed.  This assertion agrees with simple
physical intuition and is supported by the results of all-atom
molecular dynamics (MD) simulations. According to MD, the TTCs around 0.8
are observed in nearly optimal conditions.  Reaching higher values
requires special efforts and it should be difficult in practice. The TTC
can be partially responsible for the persistent discrepancies between the
twisting rigidity of DNA measured by different methods.
\end{abstract}

\pacs{87.14.gk 87.15.H- 87.15.ap 87.15.ak}

\maketitle

The twisting stiffness is a unique property of DNA that makes possible
its supercoiling, which is essential for genome compaction and
regulation \cite{Vologodskii:94b}. At present, the mechanics of
supercoiled DNA is frequently studied by using nanotechniques that
offer a means to stretch and twist single molecules
\cite{Smith:92b,Strick:96,Wang:97b}. Due to its remarkable conceptual
simplicity this method has gained broad recognition. The effects of
external twisting depend upon the torsional rigidity of DNA and the
values of applied torques. Accurate knowledge of these quantities is
necessary for interpretation of experimental data.  The recent
methodological advances made twisting of single DNA with external
torques of known magnitude \cite{Bryant:03,Forth:08} possible.
However, one question has never been asked: Is it evident that
the twisting stress created inside DNA always corresponds to the
external torque measured from outside?

\begin{figure}[ht]
\centerline{\includegraphics[width=8cm]{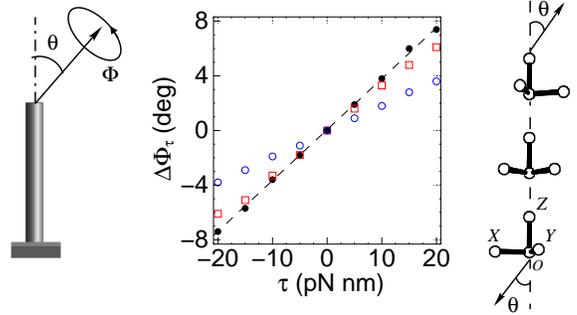}}
\caption{\label{Fbd1} (Color online) (Left panel) An elastic rod is fixed
on a solid base. A twisting torque is applied  at the top with the
torque axis inclined by angle $\theta$ form the vertical axis. 
(Right panel) A discrete WLR model of a DNA fragment of three base
pairs. The external torques are applied to terminal bases under angle
$\theta$ with respect to the chain axis. (Middle panel) Twisting
caused by external torques in BD of a 14-mer DNA fragment with
elasticity corresponding to MD data \cite{Mzpre:11}. The torques were
applied with $\theta=0^\circ$ (black dots), 30$^\circ$ (red squares),
and 60$^\circ$ (blue circles). The black dashed line displays the
theoretical dependence for the input value of $l_t$.
}\end{figure}

A simple example shown in \Rfg{bd1} demonstrates that the answer to
the above question can be negative. Consider the left panel. An
elastic rod is fixed on a solid base, with a constant torque $\tau$
applied to its top, which causes a deviation of the equilibrium twist
angle $\Delta\Phi_\tau=\Phi_\tau-\Phi_0$. For simplicity, assume that
the bending rigidity is very high and bending deformations can be
neglected. If the torque axis is deviated by angle $\theta$ only the
vertical torque projection works and the twisting stress inside the
cylinder as well as $\Delta\Phi_\tau$ decrease as $\cos\theta$. The
above effect persists with the addition of other degrees of freedom.
\Rfg{bd1} illustrates this for Brownian dynamics (BD) of a discrete
wormlike rod (WLR) model where bending, twisting and translational
motions of DNA are reproduced in agreement with experiments
\cite{Mzjpc:09}. In the harmonic approximation the
torsional part of the free energy is
\beq\label{EU}
\frac{U_\tau(\Phi)}{kT}=\frac{l_t}{2L}\left(\Phi-\Phi_\tau\right)^2
\eeq
where $L$ is the DNA length, $k$ is Boltzmann's constant, and $T$ is the
temperature. The twisting rigidity of DNA is characterized by parameter
$l_t$ called the torsional persistence length. In this case
\beq\label{EPhi}
\Delta\Phi_\tau=\frac{\tau L}{kTl_t}
\eeq
and the probability distribution of torsional fluctuations is Gaussian
\beq\label{EPphi}
P_{\Phi}\sim\exp\left[-\frac{l_t}{2L}\left(\Phi-\Phi_\tau\right)^2\right].
\eeq
The right panel of \Rfg{bd1} shows a coarse-grained WLR representation
of a DNA trimer. Each base pair is modeled by a rigid composite bead of
four particles $O$, $X$, $Y$, and $Z$ that define a local Cartesian frame. The
$OZ$ axis is constrained to pass through the center of the following bead
in the chain. Other degrees of freedom are restrained by harmonic
nearest-neighbor potentials adjusted to provide the desired macroscopic
rigidity for bending and twisting \cite{Mzjpc:09}. The twisting torques
are applied to terminal composite beads by using the earlier algorithm
\cite{Mzjctc:09}, with a small modification to deviate the torque
vector by a fixed angle $\theta$ with respect to the local $OZ$ axes.
The external torque at one end is compensated by reactions at the
opposite end so that the integral external force and torque on the
molecule are always zero. This can be viewed as if a demon in the moving
global frame twists one end of the molecule and holds the opposite end
so that the overall translation and rotation are not perturbed.  The
results of BD simulations of a 14-mer DNA fragment are shown in the
middle panel.  As expected, the effective torque decreases
approximately as $\cos\theta$.  The analytical linear dependence
corresponding to the true $l_t$ value is reproduced in BD only with
$\theta$=0.

\begin{figure}[ht]
\centerline{\includegraphics[width=6cm]{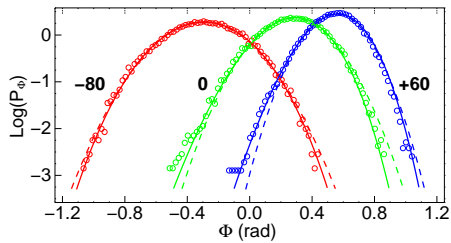}}
\caption{\label{Fpbd1} (Color online)
The normalized probability distributions of twisting fluctuations in
DNA obtained by all-atom MD simulations \cite{Mzpre:11}. The $\Phi$
angles are measured directly between the terminal base pair frames;
therefore, $\Phi=0$ for one helical turn. The CG dodecamer was considered
in relaxed state ($\tau=0$) and under steady torsional stress induced by
external torques of $\tau=$ -80 and +60 pN nm, respectively. The dashed
lines show the corresponding Gaussian distributions [\Req{Pphi}] for
the computed values of $\Phi_\tau$ and $l_t$. The solid lines display the
analytical distributions for the best-fit anharmonic torsional potential
}\end{figure}

It might seem that the above example is irrelevant because in experiments
the external torques are always applied along the stretched DNA double
helix. The problem is that local twisting axes in real DNA may be
inclined with respect to the helix direction. This issue surfaced
unexpectedly in the recent study of elastic properties of strained DNA by
all-atom molecular dynamics (MD) \cite{Mzprl:10,Mzpre:11}. In these
computations steady twisting torques were applied to two short DNA
fragments that I denote here as TA and CG (because of the AT- and
GC-alternating sequences, respectively). The classical all-atom MD
produce the probability distribution of twisting $P_\Phi$. The three
distributions  shown in \Rfg{pbd1} were obtained for dodecamer CG (one
helical turn) with contrasting values of the external torque. From these
data, the $l_t$ value can be extracted by using either \Req{Phi} or
\req{Pphi} and the resulting two values should be identical.  However,
according to \Req{Pphi} the variance of the distribution should be
constant whereas in \Rfg{pbd1} it evidently changes with $\tau$; that is,
the harmonic approximation breaks down. Therefore, a broader issue should
be considered, namely, whether or not these MD results agree with
statistical mechanics of an anharmonic elastic rod. To answer this
question we look for an anharmonic potential that would consistently
describe the MD data. Assuming that $\Phi_\tau=0$ for zero load the free
energy is
$$
U_\tau(\Phi)=-\tau\Phi+U(\Phi),
$$
where $U(\Phi)$ is the torsional potential of bare DNA, therefore,
\beq\label{Etau1}
\tau=\frac{dU}{d\Phi}\Big|_{\Phi_\tau}.
\eeq
According to \Rfg{pbd1} all the probability distributions are nearly
Gaussian; that is, in the vicinity of $\Phi_\tau$ the harmonic
approximation is valid and one can use the Taylor series to obtain
\beq\label{El_t1}
l_t(\tau)=\frac{L}{kT}U''(\Phi_\tau).
\eeq
Let us look for a polynomial form of $U(\Phi)$. We choose $U(0)$=0; in
this case the first term is quadratic. For $l_t(\tau)$ not to be
constant the polynomial should be of the third degree or higher, for
instance,
\beq\label{EU1}
U(\Phi)=q_1\frac12\Phi^2+q_2\frac16\Phi^3+q_3\frac1{24}\Phi^4.
\eeq
In this case \Req{tau1} and \req{l_t1} give
\bea\label{Etau2}
\tau&=&q_1\Phi_\tau+q_2\frac12\Phi_\tau^2+q_3\frac1{6}\Phi_\tau^3\\
\label{El_t2}
l_t&=&\frac{L}{kT}\left(q_1+q_2\Phi_\tau+q_3\frac1{2}\Phi_\tau^2\right).
\eea
MD provides $\Phi_\tau$ and $l_t$ measured for several values of
$\tau$. Coefficients $q_i$ are found by the least squares minimization
of the discrepancy between the left- and right-hand parts of
\Req{tau2} and \req{l_t2}. The results of this fitting are shown in
\Rfg{vstq1} by dashed red lines. It is seen that they
significantly deviate from the MD points. The deviations are clearly
systematic and they go far beyond the error ranges.  This situation
does not change with increased degree of the polynomial.  Note that in
all plots the deviations grow with $\left|\tau\right|$.  It turned out
that the quality of fitting is radically improved if a torque transfer
coefficient (TTC) $q_0$ is added to \Req{tau2} as follows:
\beq\label{Etau3}
q_0\tau=q_1\Phi_\tau+q_2\frac12\Phi_\tau^2+q_3\frac1{6}\Phi_\tau^3,
\eeq
with $q_0$ optimized together with other $q_i$. In this case,
the agreement within the range of statistical errors is easily reached
for polynomials of the fourth degree and higher. Interestingly, the
optimized TTC values are similar for CG and TA (0.81 and 0.79,
respectively) and they vary by only a few percent when the degree of the
polynomial is increased up to six. The quality of fitting is illustrated
by the solid traces in \Rfg{pbd1} and \rfg{vstq1}.  The agreement is
quite good, notably, small deviations of $P_\Phi$ from Gaussians in
\Rfg{pbd1} are well reproduced. These subtleties have low statistical
weight and actually do not affect the values of $\Phi_\tau$ and $l_t$
involved in the fitting. One can conclude that the MD results
qualitatively and quantitatively agree with statistical mechanics, but
only if the occurrence of the TTC $q_0$<1 in \Req{tau3} can be reasonably
explained.

\begin{figure}[ht]
\centerline{\includegraphics[width=8cm]{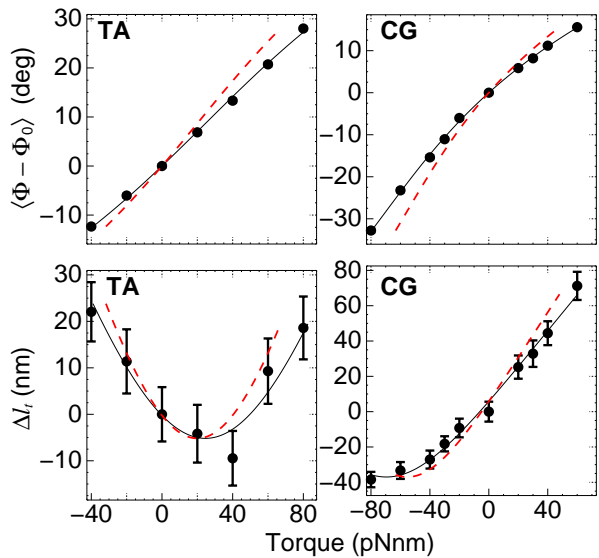}}
\caption{\label{Fvstq1} (Color online) Fitting of analytical models to
the results of all-atom MD simulations of two DNA fragments (TA and CG)
\cite{Mzpre:11}. Error bars show statistical errors. In the upper two
panels the symbol size corresponds to maximal errors. The dashed red
lines show the results of fitting for a polynomial torsional potential
without a TTC applied. The solid black lines show best fit dependences
for a fourth order polynomial with optimized TTC.
}\end{figure}

The difference between \Req{tau2}, and \req{tau3} indicates that good
fitting requires that the $q_i$ values in \Req{tau2} and \req{l_t2}
differ by a constant factor. Therefore, the scaling might be
transferred from \Req{tau3} to \Req{l_t2}, which would give equally
good fitting.  At first sight, this seems reasonable because the
scaling can be interpreted as a correction of the uncertainty in the
$L/T$ value.  Actually, however, this factor is used in computing
$l_t$ from MD data and a correcting scaling should be applied
simultaneously to both sides of \Req{l_t2}. Therefore, the occurrence
of $q_0$ can be reasonably explained only in \Req{tau3} by assuming
that a fraction of external torque applied to DNA is somehow lost.

The possible origin of incomplete torque transfer was shown in
\Rfg{bd1} above. The DNA can respond to external torques by either
twisting or bending or both. The torque fraction that causes twisting
varies with angle $\theta$ whereas the remaining part is largely
canceled by auxiliary reaction forces. Such forces are always
necessary. In the left panel of \Rfg{bd1} and in single molecule
experiments the reaction forces maintain DNA straight and fixed in
place. In computer simulations, they are used to zero the integral
external force and torque that otherwise would accelerate the overall
translation and rotation \cite{Mzjctc:09}.  All these reactions do not
affect torsional fluctuations, but effectively damp all other
deformations, which is where a part of the external torque goes.

The orientation of the twisting axis in real DNA is not known. The
external torques in MD are applied to terminal bases, with the torque
axes orthogonal to the base plane \cite{Mzjctc:09}. On closer
inspection, however, one notices that this choice may not be optimal.
There are other distinguished directions, for instance, the helical
axis or the perpendicular to the base-pair plane.  However, these
directions are well defined and coincident only in the ideal B-DNA
model. In reality, bases are not perpendicular to the helix direction
and paired bases are not co-planar, therefore, the definition of the
base pair plane as well as the helical axis is a matter of convention.
Moreover, local twisting occurs due to many rotatable covalent bonds
and it is possible that the transfer of the torsional stress to DNA
depends upon specific orientations of these bonds near the point where
the torque is applied.

To get an idea of the situation in real DNA, I tried to vary the
$\theta$ angle in MD similarly to the BD simulations presented in
\Rfg{bd1}. The protocols of MD simulations were the same as before
\cite{Mzprl:10,Mzpre:11}. The external torque is applied by
using a Cartesian frame rigidly attached to the base \cite{Mzjctc:09}.
The axes of this frame approximately correspond to the standard
convention (see \Rfg{md1}), and by default the torque is directed
along the $Z$ axis. Unlike the WLR model, the all-atom DNA does not have
cylindrical symmetry, therefore  with $\theta\ne0$, we should also
check different torque azimuths. The natural qualitatively distinct
directions of deviation correspond to $XZ$ and $YZ$ planes.  With
$\theta\ne0$ in the $XZ$ plane, the torque vector is inclined toward the
major ($\theta<0$) or minor ($\theta>0$) groove. In the $YZ$ plane, the
torque can be inclined toward the backbone of the same ($\theta<0$) or
the opposite ($\theta>0$) base. The magnitude of the external torque
was fixed at $\pm$40 pN nm. A relatively large value is required to
obtain variations beyond the range of statistical errors.  Earlier
studies showed that, in MD, the properties of short DNA fragments
change smoothly in this range of torques \cite{Mzprl:10,Mzpre:11}.

\begin{figure}[ht]
\centerline{\includegraphics[width=7cm]{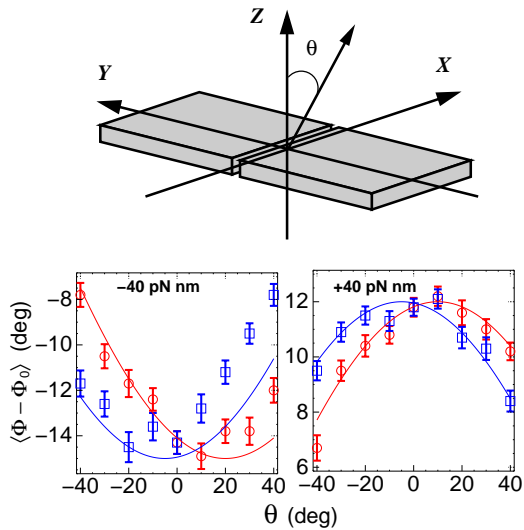}}
\caption{\label{Fmd1} (Color online)
The upper panel displays the orientation of the principal local axes
according to the standard convention \cite{Olson:01}. The $Z$ axis is
directed along the double helix. The $X$ axis looks towards the major
groove. The $Y$ axis is directed along the base pair so that the three
axes form a right-hand triple. The torque vector is inclined by a
variable angle $\theta$ with respect to the $Z$ axis. The lower panel
displays the results of all-atom MD simulations of DNA subjected to
external torques $\tau=\pm40$ pN nm applied to terminal bases. Red
open circles show the results for the torque vector inclined in the $XZ$
plane (toward the grooves), with $\theta>0$ corresponding to
deviations toward the minor groove. Similar data for the $YZ$ plane
(deviations toward the backbones) are shown by blue open squares. In
this case, the positive $\theta$ values correspond to deviations
toward the opposite base. The solid lines display the plots of
$\delta\cos(\theta-\theta_s)$, with $\delta$ and $\theta_s$ fit
manually to match the computed points.
}\end{figure}

The results shown in \Rfg{md1} confirm that the effect is also
significant in the all-atom MD.  Computed points are compared with
theoretical dependences of the form $\delta\cos(\theta-\theta_s)$, with
$\delta$ and $\theta_s$ adjusted manually. Angle $\theta_s$ specifies
the softest local twisting axis. Parameter $\delta$ affects
simultaneously the vertical displacement and the inclination of the
curves, i.e., the fitting is constrained. The agreement is relatively
good for twisting and less so for untwisting. In both planes the
softest axes are found close to, but not exactly perpendicular to the
base. For deviations in the $YZ$ plane the optimal $\theta_s$ value is
about -5$^\circ$ for both signs of the torque.  In the $XZ$ plane the
softest twisting direction is observed at $\theta_s=+10^\circ$ and
+20$^\circ$ for twisting and untwisting, respectively. Therefore, the
all-azimuth softest axis may be inclined by 20-25$^\circ$ toward the
minor groove, which would correspond to a TTC value around 0.9. Taking
into account the details of the original algorithm the TTC can be
reduced to 0.87 \cite{Mznote2}. The remaining discrepancy with the best
fit value $q_0\approx0.8$ is not negligible. Moreover, when the
twisting is measured by summing the local twist angle along the chain
\cite{Mzpre:11} the optimized TTC values decrease to 0.72 and 0.76 for
TA and CG fragments, respectively. This residual discrepancy can be
attributed to conformational fluctuations. In the ideal WLR model and
BD simulations shown in \Rfg{bd1}, the fixed angle $\theta$ is measured
between the torque and the axis of local twisting.  In contrast, in
\Rfg{md1} and MD simulations, the external torque is applied with
$\theta$ fixed with respect to the base plane whereas the local
twisting axis may fluctuate with the DNA structure. As a result, even
in an optimal orientation, the torque is not strictly collinear with
the twisting axis, and this should further reduce the TTC.

The foregoing results prove that it is difficult, if ever possible, to
transfer the desired value of external torque to real double helical
DNA. This conclusion should also be valid for nanomanipulations with
isolated DNA molecules. The samples used in experiments commonly
involve a few kilobase pairs of nearly random DNA. Earlier studies
indicate that the WLR model is valid for any integral number of
helical turns starting from one \cite{Mzprl:07,Mzjpc:09}; therefore,
one helical turn used in MD already involves all essential properties
and the conclusions obtained should be invariant to length scaling. In
experiments the double helix is attached to macroscopic beads and
extended by a stretching force.  The torques are applied to the beads
in different ways and transferred to DNA through a number of point
contacts.  However, any multi-point interaction can be decomposed to
individual components, each characterized by a specific TTC. The
quasi-randomness of natural DNA sequences and the possible
sequence dependence cannot smooth out the effect because the TTC never
exceeds 1. Therefore, the induced twisting is always downgraded.

The TTC can explain some discrepancies in the recent experimental
measurements of the torsional rigidity of DNA. The value of $l_t$ (or
$C=kTl_t$) has been measured since the 1970s \cite{Hagerman:88} because it
plays a fundamental role in all manifestations of DNA supercoiling.
Before the recent single-molecule studies, $C\approx310$ pN nm$^2$ was
considered as a consensus value for random sequence DNA. This value
reasonably agrees with some indirect estimates that used single-molecule
data \cite{Vologodskii:97,Bouchiat:98}, but it is at least 20-30\% lower
than in direct measurements using forced DNA twisting by calibrated
external torques \cite{Bryant:03,Forth:08}. Several pitfalls possibly
responsible for such a difference have been discussed in the literature
\cite{Bryant:03,Fujimoto:06}. Assuming the TTC values around 0.8, which
may slightly vary between different experimental installations, all
these data can be accounted for without additional assumptions. Only one
observation cannot be explained, namely, the very high apparent
torsional rigidity ($C=440$ pN nm$^2$) reported for equilibrium thermal
torsional fluctuations of single stretched DNA \cite{Bryant:03}. I
believe that some other factors could affect this result and it
requires further consideration.

In summary, the present paper shows that because of the structural
details of the double helix, it is not possible to transfer the entire
magnitudes of external torques to the internal twisting stress of DNA,
and that a reducing torque transfer coefficient should always be assumed.
This assertion agrees with simple physical intuition and it is supported
by the results of all-atom MD simulations. Theoretical estimates indicate
that the TTCs around 0.8 should be observed for torques nearly parallel
to the double helix. Reaching higher values requires special efforts that
should take into account the details of the DNA structure and it should
be very difficult for experiments with long molecules. The TTC explains
some controversies in the experimental values of DNA twisting rigidity
measured by different methods.

\begin{acknowledgments} The author is grateful to Alex Vologodskii for
useful discussions and comments to the paper. I also thank Zev Bryant for
correspondence and valuable clarifications concerning the rotor bead
tracking method.
\end{acknowledgments}

\bibliography{mzpaper}

\end{document}